\documentclass[man,floatsintext,12pt]{apa7}
\usepackage[utf8]{inputenc}
\usepackage{rotating}
\usepackage{lineno}
\usepackage[style=apa,backend=biber]{biblatex}
\usepackage{makecell}

\title{Auditing citation polarization during the early COVID-19 pandemic}
\shorttitle{}

\authorsnames[1,2,2,3]{Taekho You,June Young Lee, Jinseo Park$^{*}$, Jinhyuk Yun$^{*}$}
\authorsaffiliations{{Institute for Social Data Science, Pohang University of Science and Technology, Republic of Korea},{Center for Global R\&D Data Analysis, Korea Institute of Science and Technology Information, Republic of Korea},{School of AI Convergence, Soongsil University, Republic of Korea}}

\authornote{\addORCIDlink{Taekho You}{0000-0002-4406-5514}

\addORCIDlink{Jinseo Park}{0000-0003-1899-1861}

\addORCIDlink{June Young Lee}{0000-0001-8606-2429}

\addORCIDlink{Jinhyuk Yun}{0000-0002-0218-1699}

$^{*}$ Corresponding author(s). Email: jayoujin@kisti.re.kr, jinhyuk.yun@ssu.ac.kr}

\abstract{The recent pandemic stimulated scientists to publish a significant amount of research that created a surge of citations of COVID-19-related publications in a short time, leading to an abrupt inflation of the journal impact factor (IF). By auditing the complete set of COVID-19-related publications in the Web of Science, we reveal here that COVID-19-related research worsened the polarization of academic journals: the IF before the pandemic was proportional to the increment of IF, which had the effect of increasing inequality while retaining the journal rankings. We also found that the most highly cited studies related to COVID-19 were published in prestigious journals at the onset of the epidemic. Through the present quantitative investigation, our findings caution against the belief that quantitative metrics, particularly IF, can indicate the significance of individual papers. Rather, such metrics reflect the social attention given to a particular study.}


\begin{document}

% Use letters for affiliations, numbers to show equal authorship (if applicable) and to indicate the corresponding author

\maketitle

\section{Introduction}

%COVID-19 pandemic을 해소하기 위해 많은 연구자들이 COVID-19 research work 을 발표하였다. COVID-19 paper의 증가는 
The recent pandemic has boosted COVID-19-related research, which has led to a growing number of researchers publishing COVID-19-related publications~\parencite{ioannidis2022massive}. During the pandemic, as of 2021 more than 4\% of published research papers focused on COVID-19~\parencite{ioannidis2022massive}. The availability of COVID-19-related research has supported the public to overcome the current pandemic. 

% COVID-19 paper의 증가는 IF의 큰 변화로 이어졌다.
The expansion of this new research field has had a substantial impact on the scholarly publishing ecosystem. COVID-19-related publications received a large number of citations in a short period, causing a dramatic shift in citation counts. Specifically, some journals benefited from publishing COVID-19-related research because it significantly increased their mean citation rate. As an illustrative example, the \textit{Lancet} more than doubled its impact factor (IF) from $79.323$ to $202.731$, according to the 2021 Journal Citation Reports (JCR) released in June 2022. It has been contended that COVID-19-related publications have inflated the citation-based metrics; indeed, some journals have increased their IF by more than tenfold~\parencite{JCRreport}.

%이 현상은 IF의 debate에 다시한번 불을 지폈다
Consequently, the long-lasting IF controversy has reemerged. Due to the heavy-tailed nature of citation, which is sometimes referred to as the rich-get-richer effect, many critics argue that IFs do not accurately reflect the impact of scientific items because they rely solely upon mean citation counts~\parencite{pendlebury2009use,lariviere2016simple}. In response, alternative metrics have been proposed~\parencite{moed2010measuring,bradshaw2016rank}. While journal-level citation metric is widely used for assessing the impact of the research output, it has been challenged that journal-level analysis cannot evaluate the quality of individual studies. Thus, individual article-level metrics have also been introduced. One straightforward metric is the raw citation count, yet it is hard to compare the raw citation between two papers published in different fields due to differences in publication and citation cultures; various field-normalized indicators, therefore, are naturally introduced. \textcite{waltman2011towards} suggested a mean-normalized citation score (MNCS) with field-normalization, in response to the criticism of crown indicator, which is computed by dividing the number of citations in a given work by the average number of citations in papers published in the same topic area and publication year relying on the citation databases (e.g., Web of Science)~\parencite{gingras2011there}. However, one criticized that the citation distribution is skewed, so the average may not be a representative value (or the expected value) of citations for each field. When a paper is published in more than one field according to the bibliographic databases, there is one more layer of complexity. To overcome the limitation, the Relative Citation Rate (RCR), a new field-normalized effect indicator, was also proposed. The design of the indicator is similar to the MNCS, yet the expected value is calculated based on co-cited papers~\parencite{hutchins2016relative}. Still, instability might result from the limited number of co-cited papers. The Citation score normalized by cited references (CSNCR), which citations of a target paper divided by the average number of cited references in a subject area, can be an alternative to such field-normalized indicators; the indicator possesses homogeneous and consistency normalization as suggested by \textcite{bornmann2016citation}. On the other side, the distribution of citation data is usually very skewed with only a few papers being highly cited~\parencite{bornmann2013use}. However, such methods suffer from the same limitation: the values can change greatly depending on the selection of the comparison group. Moreover, although the IF metric was designed to measure the performance of journals rather than single papers~\parencite{garfield1972citation}, it is nevertheless frequently misunderstood to reflect the quality of an individual paper~\parencite{lozano2012weakening}. The spreading of these misunderstandings has increased unintended dynamics in the conduction and evaluation of research~\parencite{calcagno2012flows,rafols2012journal}, even leading to cases of malpractice~\parencite{you2022disturbance}.

Resolving this IF controversy from COVID-19-related publications necessitates a deep comprehension of citation dynamics in academia, such as the extent to which COVID-19 publications affect journal IFs and who benefits more from publishing COVID-19-related publications. The Matthew effect~\parencite{merton1968matthew}, also known as the rich-get-richer effect, gives valuable insight into the accumulation of rewards in academia~\parencite{perc2014matthew,bol2018matthew,nielsen2021global,huang2020historical,lawson2021funding}. Previous studies demonstrated that a little variation in early stages leads to a substantial difference in the productivity and citations of authors and journals in later stages~\parencite{petersen2011quantitative,bol2018matthew,nielsen2021global}. Moreover, a paper is more likely to receive citations when published in a prestigious journal that has a high IF~\parencite{lariviere2010impact}. Citation inequality results from the widening gaps in return from such small, initial differences~\parencite{allison1982cumulative,van2014field,lawson2021funding}. We believe that the emergence of the COVID-19 research field presents an excellent opportunity to comprehend scholarly dynamics in response to external societal influence. In addition, since the COVID-19 research field emerged in a very short time period, observing the COVID-19-related publications can provide a great opportunity to observe the effect of IF on the citation dynamics.

In this study, we quantitatively exhibit the impact of COVID-19-related publications on the citation ecosystem to aid in resolving the long-lasting debates on the IF metric. For this purpose, we investigate the changes in IF by publishing COVID-19-related publications considering journal IF. Our investigation builds on earlier studies on publication and citation trends during COVID-19. As former studies demonstrated that COVID-19-related publications received more citations than other publications ~\parencite{pajic2023covid,ioannidis2022massive,zheng2023significant}, we attempt to investigate the hypothesis that these citations are the result of great attention from other fields. We also investigated the relationship between the COVID-19-related publications and the IF to extend the results that the COVID-19-related publications received a larger number of citations~\parencite{pajic2023covid,ioannidis2022massive,zheng2023significant}; we assessed whether the surplus IF is correlated with the number of COVID-19-related publications. To fully understand the impact and potential future trends of academia, we finally examined the publication dynamics of highly-cited COVID-19-related publications by analyzing the journals, the date of publication, and their influence on the citation ecosystem.

\section{Methods}\label{sec:method}

\subsection{Data}
We used publications and citation data from the \texttt{XML} dump of the Web of Science Core Collection, which is dated back to 2017 and updated until the 26th week of 2022. The data includes complete copies of Science Citation Index Expanded (SCIE), Social Sciences Citation Index (SSCI), and Arts \& Humanities Citation Index (AHCI), along with the Emerging Sources Citation Index (ESCI). The data comprises 16,957,120 articles, 82,317 journals, and 116,086,223 references retrieved from papers published between 2017 and 2022.

\subsection{COVID-19-related publications}\label{COVID-19-related}
COVID-19-related publications were retrieved from the Web of Science database (WOS, url{https://www.webofscience.com/}) using the following search queries provided by Dimensions (\url{https://www.dimensions.ai/covid19/}): \texttt{"2019-nCoV" OR "COVID-19" OR “SARS-CoV-2” OR "HCoV-2019" OR "hcov" OR "NCOVID-19" OR "severe acute respiratory syndrome coronavirus 2" OR "severe acute respiratory syndrome corona virus 2" OR “coronavirus disease 2019” OR [("coronavirus" OR "corona virus") AND (Wuhan OR China OR novel)]}. We limit the publications issued from the first day of 2019 because the search query includes \textit{"coronavirus"}, which finds publications regardless of year. Before 2019, or before the COVID-19 pandemic, publications ought to consider different coronaviruses. A total of $251,718$ COVID-19-related publications were collected on 4 July 2022.  
Note that the query \texttt{"("coronavirus" OR "corona virus") AND (Wuhan OR China OR novel)"} was included because some publications in the early stages of the pandemic were only found using these terms, particularly published before its official naming by WHO on February 11, 2020; we found 461 publications containing these terms, and 298 publications are published before 2021. These publications received 197.9 citations on average, with 13,713 maximum citations suggesting their impact. Thus, even though these terms could be politically controversial, we include them. We consider all other papers in the WOS that were not retrieved from the above searching process as non-COVID-19-related publications.

\subsection{Estimation of the power-law exponent}
It is commonly observed that the distribution of citations has heavy tails, such as power-law or lognormal distributions~\parencite{radicchi2008universality,eom2011characterizing}, yet it is hard to generalize. However, extreme values can occur in heavy-tailed distributions, regardless of whether the distribution is lognormal or power-law. In this study, we use power law exponent $\alpha$ a measure of skewness: When the exponent $\alpha$ of the power-law distribution is high, a few highly-cited papers exist, while relatively more highly-cited papers exist when the exponent is low~\parencite{krapivsky2000connectivity}. Therefore, the power-law exponent can show the proportion of highly-cited papers, i.e., how much the citations are skewed.

The primary generative mechanism of the power-law distribution is often attributed as a preferential attachment~\parencite{barabasi1999emergence} that a paper's likelihood of being cited increases with the number of citations it receives, which is naturally linked to the Matthew, or rich get richer, effect. Previous research ~\parencite{price1976general,golosovsky2012stochastic,petersen2014reputation,yin2017time} proposed the model of the citation dynamics, but the underlying mechanism is not fully understood. In this study, the power-law exponents of the citation distribution in Fig.~\ref{fig:fig1}A were estimated using the Python package named \texttt{powerlaw}~\parencite{alstott2014powerlaw}. Although all the citation distributions in Fig.~\ref{fig:fig1}A seem to be heavy-tailed distributions, which are commonly referred to as the power law, we verified that the distributions sincerely follow the power law via comparison with alternative distributions (e.g., log-normal or exponential). In the comparison with the exponential distribution, all distributions were found to be more likely to be power-law distributions rather than exponential ($p<0.001$). However, the comparison between log-normal distribution and power-law distribution was not statistically significant, i.e., it is unclear which distribution provides the best fit. Only non-COVID-19-related publications published in 2021 better fit the power-law distribution in a statistically significant manner, while the other three were inconclusive ($p$ varied $0.48$--$0.60$). In this study, we estimated the power-law exponent with the assumption of a simple power law ($y \sim x^k$) regardless of the best fit distribution, as we were more interested in comparing the thickness of the tails than in determining the exact exponents. 

To fit the distribution, we applied a simple linear regression method to the logarithm of the values of interest to estimate the power-law scaling relationship between the IF and its surplus by COVID-19-related publications, assuming a simple power-law scaling of $y = Cx^k$.

\subsection{Reproduction of the journal impact factors}
Although we extracted the total number of publications in the WOS with a complete copy of the WOS provided by Clarivate, minor differences can be presented mainly because the WOS does not report detailed methods to filter the dataset, e.g., dump dates and the coverage of citable items. Thus, to reproduce and estimate the journal impact factors (IFs), we followed the method used for the JCR impact factor~\parencite{JCRreport} but with an in-house \texttt{XML} copy of the Web of Science, as follows:

\begin{equation}\label{eq:impact}
\mbox{IF} = \frac{\mbox{citations received by items in the given year}}{\mbox{number of citable items published in the past 2 years}}.
\end{equation}

\noindent We limited the citable items to those belonging to the journals indexed in SCI-Expanded, SSCI, and A\&HCI. We also considered as citable items only articles, review papers, and proceedings papers in terms of publication type; however, publication types were not considered when computing the number of citations.

Note that, as of 2020, Clarivate Inc. now considers early access publications as regular publications and includes them in the calculation of IF. For instance, if an article is published as early access in 2020 and officially published in 2021, then the article is counted as a citable item published in 2020, taking into account the references as the citations occurred in 2020. The article is not considered in 2021.

As an illustrative example, \textit{CA-A CANCER JOURNAL FOR CLINICIANS} published 61 publications in 2019 and 2020, which received 15,037 citations from those published in 2021. These publications include 37 articles, 13 editorial materials, 9 reviews, 1 letter, and 1 correction. The journal has 53 citable items published in 2019 and 2020, which consist of 13 review papers and 40 articles. From the data, we calculated the IF of the journal in 2021 as 283.72. We found 3 COVID-19-related publications as citable items in 2020, and they received 19 citations. When we exclude these publications and their citations, the IF increases to 300.36.

With this procedure, we successfully reproduced IF scores that highly correlated with the IFs provided by Clarivate JCR (Spearman $\rho =0.99$; see Fig.~\ref{fig:sm_IF}). In this study, we refer to the value computed from Eq.~\ref{eq:impact} as IF instead of the impact factor provided by JCR unless otherwise specified. When computing the IFs excluding COVID-19-related publications, we counted out the COVID-19-related citable items and their received citations from the denominator and numerator in Eq.~\ref{eq:impact}, respectively. 

\subsection{Keyword co-occurrence analysis}

Since defining the research field to compare the citations between COVID-19-related works is difficult, we employed keywords of each paper as a proxy for the similarity between the two papers. Therefore, we randomly selected papers as a control group for COVID-19-related publications based on the research category and publication year; and then compared the keyword-based similarity between papers in the citation relationship.

In detail, first, we compute the number of COVID-19-related publications in each research category and publication year. Second, we randomly selected the same number of papers while keeping the research category and publication year. We repeated 20 times for the selection process to check the robustness of the random selection. Third, we compared the keyword co-occurrence ratio between the target paper and its references or citations for three types: (1) COVID-19-related publications and their all references or citations, (2) COVID-19-related publications and the references or citations that are also COVID-19-related publications, and (3) randomly selected publications and their all references or citations. We counted the number of references or citations sharing at least one keyword. We also calculated the Jaccard similarity of keywords between the focal publication and each reference or citation of the publication and calculated the average Jaccard similarity.

\begin{figure}[btp]
    \centering
    \includegraphics[width=0.8\linewidth]{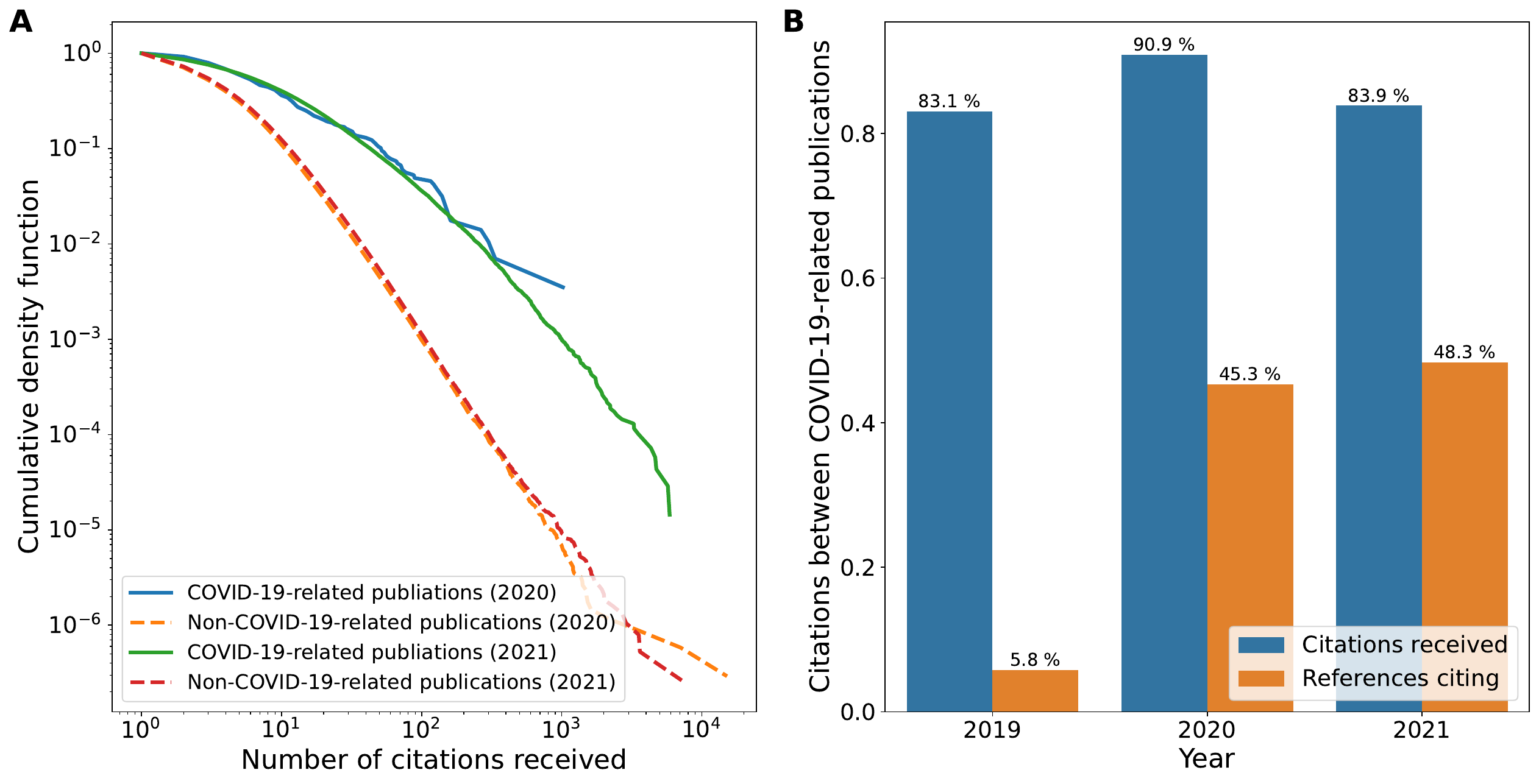}
    \caption{\textbf{Difference in citation distribution between COVID-19-related and non-COVID-19-related publications.}
    \textbf{A} Citation distribution of COVID-19-related and non-COVID-19-related publications that contribute to the annual IF calculation.
    \textbf{B} Citation origin of COVID-19-related publications. We display both the percentage of citations received from other COVID-19-related publications and references citing other COVID-19 publications.}
    \label{fig:fig1}
\end{figure}

\section{Results}
\subsection{Citation exchange between COVID-19-related publications}

During the pandemic, COVID-19-related publications have increased their share in academia. In 2019, only 350 papers (0.013\%) were related to COVID-19, many of which were mainly focused on other coronaviruses, based on our search query (see Methods for step-by-step details on gathering COVID-19-related publications). As the virus spread, their share increased to 89,112 (2.004\%) and 162,256 (4.194\%) in 2020 and 2021, respectively. Moreover, COVID-19-related research occupied a major fraction of all citations across academia. Such papers published in 2020 received 2,654,613 citations until the end of 2021, which is 13.8\% among the total 19,203,421 citations in 2020 and 2021. But not only gaining a high share, COVID-19-related publications also received immediate citations: those published in 2021 received 787,009 citations out of the total 6,457,473 citations in 2021 (12.2\%). This same trend even extended down to the monthly citation level, as displayed in Fig.~\ref{fig:sm_cite_month}. After publication, 31.8\% of the citations of COVID-19-related publications arose within 6 months, while 22.2\% did so for non-COVID-19 papers. Compared to the statistics indicating that COVID-19-related publications produced just 4.1\% and 6.9\% of references within the same time period (2020 and 2021, respectively), this proportion of received citations is high.

The increased attention given to COVID-19 research resulted in a citation distribution with a longer tail than other research. The two-year citation distribution shows that COVID-19-related publications received more citations than non-COVID-19-related publications in a given year (see Fig.~\ref{fig:fig1}A for the merged distribution along with Fig.~\ref{fig:sm_citation_distribution} displaying separated distributions). Note that the merged distributions considered the citations used to compute the IF, which shows how these publications can contribute to increasing IF. For instance, in 2021, the distribution contains publications published in 2019 and 2020, and their citations received in 2021 are considered. Note that we also displayed (yearly) separate distributions in Fig.~\ref{fig:sm_citation_distribution}. We hypothesize that if the distributions of COVID-19-related publications and non-COVID-19-related publications are similar, COVID-19-related publications would simply be a part of the natural academic citation patterns. However, COVID-19-related publications and non-COVID-19-related publications show significantly different distributions (two-sample Kolmogorov-Smirnov test $p < 0.001$ for both 2020 and 2021). When we assume that the citation distribution follows a simple power law ($y \sim x^k$), the COVID-19-related publications show $k \simeq 1.9$ and $k \simeq 2.7$ for 2020 and 2021, respectively (see Methods for the detailed computation), while non-COVID-19-related publications have an exponent of $3.2$ and $3.3$ for 2020 and 2021, respectively. The lower exponents indicate that the proportion of COVID-19-related publications with extremely high citation counts is greater than that of non-COVID-19-related publications. Indeed, we found that 225 COVID-19-related publications (0.3\% of all COVID-19-related publications in 2021) received more than 500 citations, in 2021, while only 132 non-COVID-19-related publications (0.003\% of all non-COVID-19-related publications in 2021) received more than 500 citations. Consequently, COVID-19-related publications also received more citations on average. COVID-19-related research received an average of $22.6$ (2020) and $21.8$ (2021) citations, while non-COVID-19-related publications received $4.9$ (2020) and $5.2$ (2021) citations. This result is consistent with a previous observation using SCOPUS~\parencite{ioannidis2022massive}.

We found that the high citation counts of COVID-19-related publications mostly come from other COVID-19-related publications. In Fig.~\ref{fig:fig1}B, more than 80\% of citations that COVID-19-related publications received come from other COVID-19-related publications. We also observed that more than 40\% of the references that COVID-19-related publications produce are heading to COVID-19-related publications, excluding 2019.

Since most COVID-19-related publications have been published in a short period from diverse fields, their citations can be more diverse than others. To check the diversity and homogeneity of citation relations, we compare the keywords between two publications in a citation relationship (see Methods). We also randomly sampled the same number of publications with corresponding COVID-19-related publications from the same category and publication year as the control set. First, COVID-19-related publications share the same keywords with 25.6\% of their references, which is similar to samples ($\sigma=0.00043$). However, if the references are also COVID-19-related publications, only 11.5\% of references have the same keywords as the focal publication(Fig.~\ref{fig:sm_keyword}). For random samples, 35.1\% ($\sigma=0.00016$) of forward citation relations share the same keywords, while only 11.6\% of forward citation relations of COVID-19-related publications share the same keywords. Second, the average Jaccard similarity between COVID-19-related publications and their references is 0.030. When we limit references to COVID-19-related publications only, the average Jaccard similarity is also 0.028. For the COVID-19-related publications and their forward citations, the average similarity is 0.015. The average Jaccard similarity of non-COVID-19-related publications is 0.028 ($\sigma = 0.00004$) with references and 0.035 ($\sigma=0.00016$) with forward citations. In short, the keyword similarity between COVID-19-related papers is low, but they cite other non-COVID-19-related publications having similar keywords. The result implies that COVID-19-related publications cite other COVID-19-related publications more, even though their topical similarity is low. In addition, the average Jaccard similarity between COVID-19-related publications and their citation counts negatively correlates (Pearson $r = -0.521$ and Spearman $\rho = -0.521$, see Fig.~\ref{fig:sm_keyword_citation}. This is also true for the IF and average Jaccard similarity (Pearson $r=-0.521$ and Spearman $\rho=-0.521$). The result implies that the more the COVID-19-related publication is cited by others, the more publications on diverse topics are citing the publication.

\subsection{Contribution of COVID-19-related research to IF inflation}

\begin{figure}[btp]
    \centering
    \includegraphics[width=\linewidth]{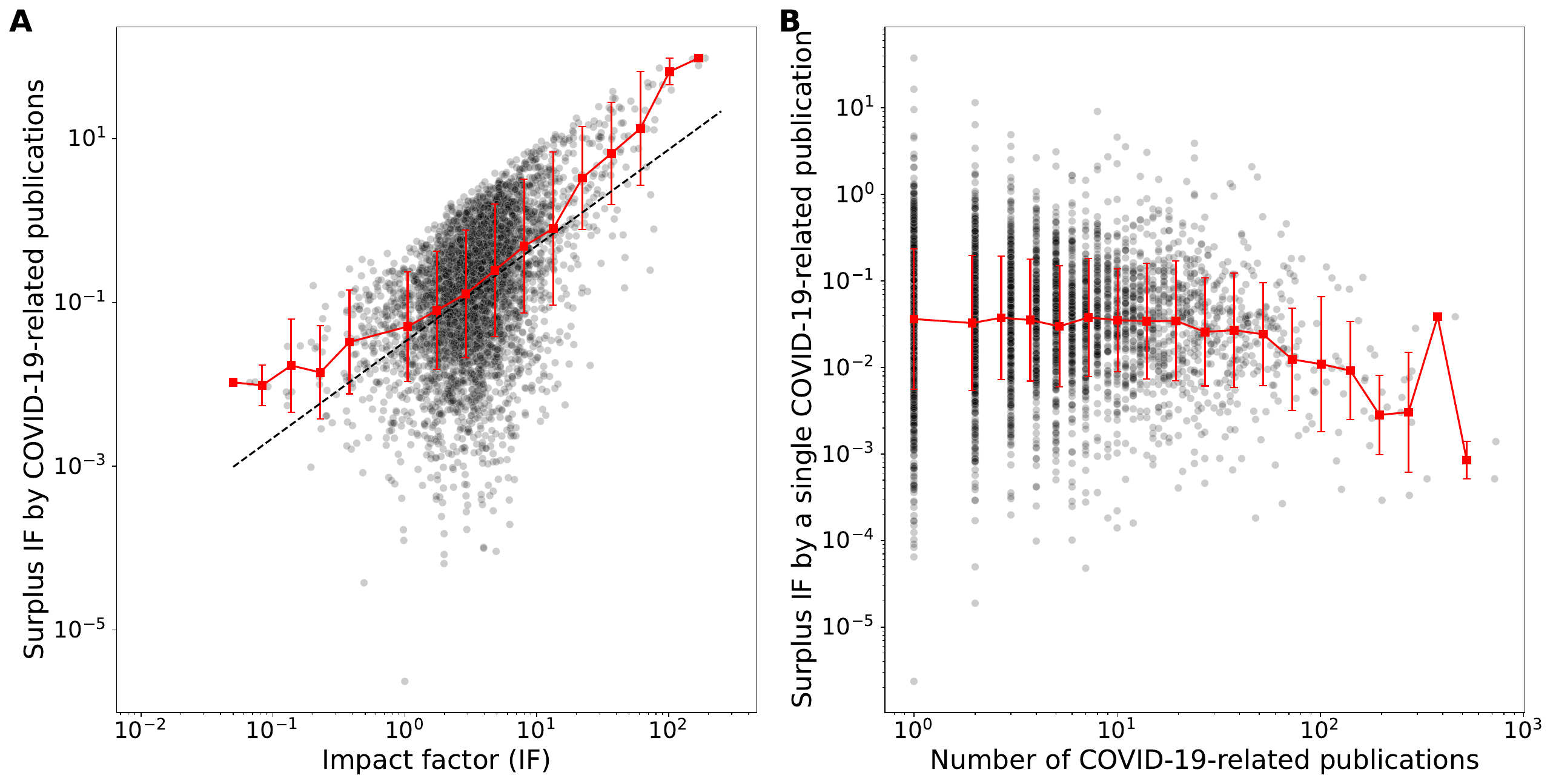}
    \caption{\textbf{Surplus impact factor (IF) by COVID-19-related publications.} 
    \textbf{A} Journal impact factor increase by publishing COVID-19-related publications, where the simple superlinear growth $y \sim x^{1.7}$ can characterize the growth pattern (dotted line). 
    \textbf{B} Increase in IF per COVID-19-related publication in proportion to the number of COVID-19-related publications published in journals. 
    In both A and B, the red dots represent the average values in the log-scale (i.e. geometric means) of surplus IF, and the error bars show the standard deviation in the log-scale.
    }
    \label{fig:fig2}
\end{figure}

The former finding suggests that several highly cited COVID-19-related publications may bolster the journals' IF. But do they increase the IFs? Because many countries use the IF as the barometer of research evaluation. To quantify the change in IF from publishing COVID-19-related publications, we calculate the IF in terms of the existence and number of COVID-19-related publications (see Methods for IF calculation). We measure two different types of IFs and compare them to estimate the advantage of publishing COVID-19-related publications: IF excluding COVID-19-related publications and IF including them.
We first compute surplus IF by differentiating the two IFs: with or without COVID-19-related publications. 
\begin{equation}
    \mbox{surplus IF} = \mbox{IF}_{\mbox{w/ COVID-19-related publications}} - \mbox{IF}_{\mbox{w/o COVID-19-related publications}}.
\end{equation}
\noindent For instance, the IF of the \textit{Lancet} in 2021 is 93.04 without any COVID-19-related publications, while the IF is 189.25 with COVID-19-related publications. Thus, it suggests the journal increased its IF by 96.21 due to COVID-19-related publications. One should note that this increase is not a yearly change but a comparison between the original IF (with COVID-19-related publications similar to the original JCR) and adjusted IF without COVID-19-related publications for the same journal in the same year. We observe that only 763 journals (16\%) among those publishing one or more COVID-19-related publications in 2019 and 2020 dropped in IF in 2021, while the other 4,004 journals (84\%) enhanced their IFs through the publication of COVID-19-related publications in the same period.  For the former, even though the journals decreased in IF by publishing COVID-19-related publications, the amount of decrease was limited. Only one of these 763 journals (\textit{CA-A CANCER JOURNAL FOR CLINICIANS}) dropped in IF by more than $1$ (Fig.~\ref{fig:sm_if_change}).

The motivations for citations vary by author and by project~\parencite{bornmann2008citation}. Proposed as a normative theory, and ideally, individual scientists cite other works because they are under the influence of the preceding scientific work~\parencite{merton1973sociology}. However, the rapid growth in the number of academic publications makes it difficult to find all related works, and thus, citation dynamics are also affected by other factors: former citations, prestige, visibility, nationality, etc~\parencite{cozzens1985comparing,wang2014unpacking,petersen2014reputation,gomez2022leading}. Another study also found that this kind of behavior is also affected by the interdisciplinarity and field of study~\parencite{lariviere2010impact}. For the COVID-19-related publications, we find that the surplus IF is proportional to the IF (Fig.~\ref{fig:fig2}). High correlation exists between IF and its surplus (Spearman $\rho=0.418$, Fig.~\ref{fig:fig2}A), and their relationship is even superlinear ($y \sim x^{1.7}$, $R^2 = 0.505$, $p<.001$). In other words, the surplus IF of a journal is much higher when the journal's IF is high. For example, if the IF of journal A is twice as high as that of journal B, on average, the surplus IF of journal A is 3.24 (=$2^{1.7}$) times more than journal B, owing to the COVID-19-related publications. This pattern is also verified when we consider the relative surplus IFs by dividing the surplus IF by the journal IF, which also shows a positive correlation (Fig.~\ref{fig:sm_IF_increase_relative}).

Since the average citation of COVID-19-related publications is high, one can insist that the surplus IF is proportional to the number of COVID-19-related publications. However, the additional increase in COVID-19-related publications did not lead to the same proportional increase in the journal IF. The comparison between the journal's number of COVID-19-related publications and the surplus IF per a COVID-19-related publication shows a diminishing return (Fig.~\ref{fig:fig2}B). Journals that published only one COVID-19-related publication in 2019 and 2020 increased their IF by $0.12$ on average, whereas journals that published over $500$ COVID-19-related publications in the same period increased their IF by only $0.0009$. For example, the \textit{Lancet}, the journal with the highest IF in JCR 2021, doubled its IF (from 93.04 to 189.25) while publishing only $46$ COVID-19-related publications (9.4\% of all citable items) in 2019 and 2020. To take one more extreme example, one journal that published only one COVID-19-related publication in 2020 increased its IF by $37$, whereas the journal that published the largest number of COVID-19-related publications (730 publications) improved its IF by only $1.02$. These examples show the limitations of using IF: the former journal only published a small number of publications, so its IF has fluctuated more by publishing COVID-19-related publications. The latter journal exhibits that publishing a large number of COVID-19-related publications did not gain many benefits. Although allocating more shares to COVID-19-related publications correlates positively with the rise in the IFs of journals, the correlation is slight (Spearman $\rho = 0.008$; see Fig.~\ref{fig:sm_covid_IF_correlation}).

To confirm that COVID-19-related research has legitimately increased journal IFs, we examine the correlation between IFs across the year accounting for the existence of COVID-19-related publications. First, the correlations between IFs of two consecutive years are high when excluding COVID-19-related publications (Pearson $r = 0.957$, Spearman $\rho = 0.925$ between 2019 and 2020, $r = 0.925$, $\rho = 0.955$ between 2020 and 2021). With COVID-19-related publications, the Pearson correlations between two consecutive years decreased ($r = 0.850$), while the Spearman correlation remained at a similar level ($\rho = 0.946$) between 2020 and 2021. Thus, the rank of journals is quite stable. The decrease in Pearson correlation demonstrates that COVID-19-related publications increased inequality in IF, and the impact of other external factors, such as random changes, was limited. If other external factors changed the IF more than COVID-19-related publications, the Spearman correlation was also decreased. We also observed a decrease in Pearson correlation ($r = 0.849$) when comparing 2020 IF without COVID-19-related publications and 2021 IF with COVID-19-related publications, and $r = 0.926$ when comparing 2020 IF with COVID-19-related publications and 2021 IF with COVID-19-related publications. However, the rank in both cases was stable (Spearman $\rho = 0.946$). In short, given that journals with a higher IF received a greater increase in IF from COVID-19-related publications (Fig.~\ref{fig:fig2}A), the publication of COVID-19 research contributes to the polarization of journal IFs.

\begin{figure}[btp]
    \centering
    \includegraphics[width=0.5\linewidth]{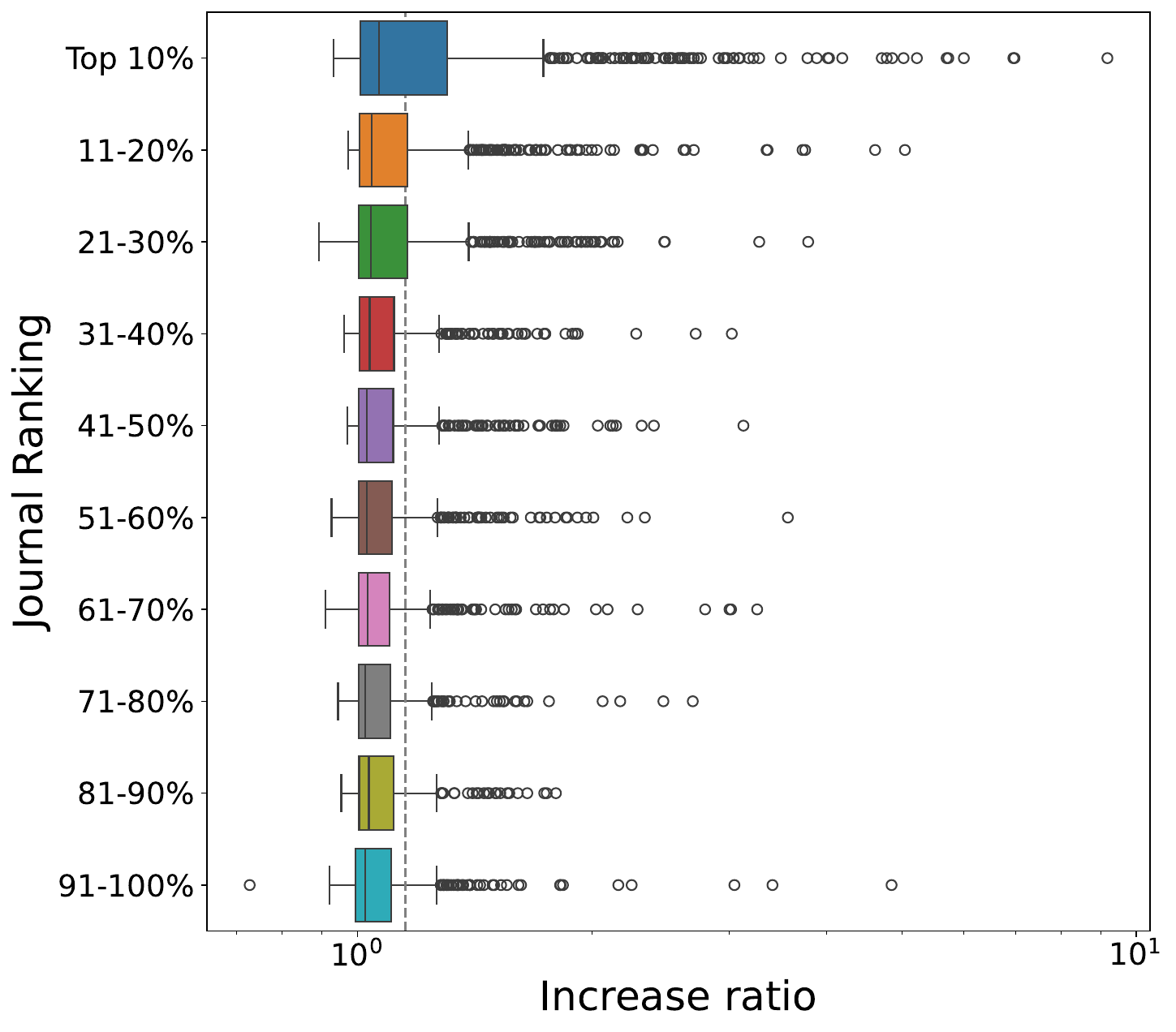}
    \caption{\textbf{Relative ratio of surplus IF from publishing COVID-19-related publications by the 2021 journal rankings for JCR categories.} The ratio was calculated by dividing the IF including COVID-19-related publications by the IF excluding COVID-19-related publications for 2021. All journals that published at least one COVID-19-related publication are accounted for, regardless of whether the IF is gained or dropped by publishing COVID-19-related publications. The dotted line indicates the average ratio from publishing COVID-19-related research (15.2\%). Here, the boxes represent the quartiles of the dataset except for points determined to be outliers.}
    \label{fig:fig3}
\end{figure}

\subsection{The Matthew effect of IF polarization during the pandemic}

In the preceding sections, we demonstrated that the publication of COVID-19-related research had a positive correlation with journal IFs, while the amount of increment had a strong correlation with the IFs of the journals (Fig.~\ref{fig:fig2}). One may wonder how much the overall journal landscape, i.e., the journal rankings, has changed due to the surplus IFs, or conversely, the magnitude of the change in IF based on the journal ranking \parencite{quaderijournal}. To demonstrate the influence of COVID-19-related publications on the landscape of JCR rankings, we compare the ratio of surplus IF in 2021 considering the journal rank in their research categories. The ratio was calculated by dividing the IF including COVID-19-related publications by the IF excluding COVID-19-related publications. On average, the publication of COVID-19-related publications increased the journal IF by 15.2\% (dotted line in Fig.~\ref{fig:fig3}). The IFs of the top 10\% prestige journals increased by 39.4\%, while the IFs of the bottom 10\% journals increased by only 9.6\% on average. Most journals increased their IF by less than the average increase (15.2\%) except for the top 20\%. On average, higher-ranked journals gained more citations, and this trend is robust across all categories (Table~\ref{tab:field}).

The majority of journals with a significant increase in IF due to COVID-19-related publications were already high-IF journals. Among the 4,767 journals that published one or more COVID-19-related publications, 132 (2.77\%) journals increased their IF by greater than twofold. 55.3\% of these 132 journals are in the top 10\% of at least one of their research categories. Only five journals fall within the bottom 10\%. In terms of research category, 102 of the 132 journals (77.3\%) are classified into \textit{Clinical Medicine} since the majority of COVID-19-related publications (70.9\%) were published in this category group.

\begin{figure}[btp]
    \centering
    \includegraphics[width=\linewidth]{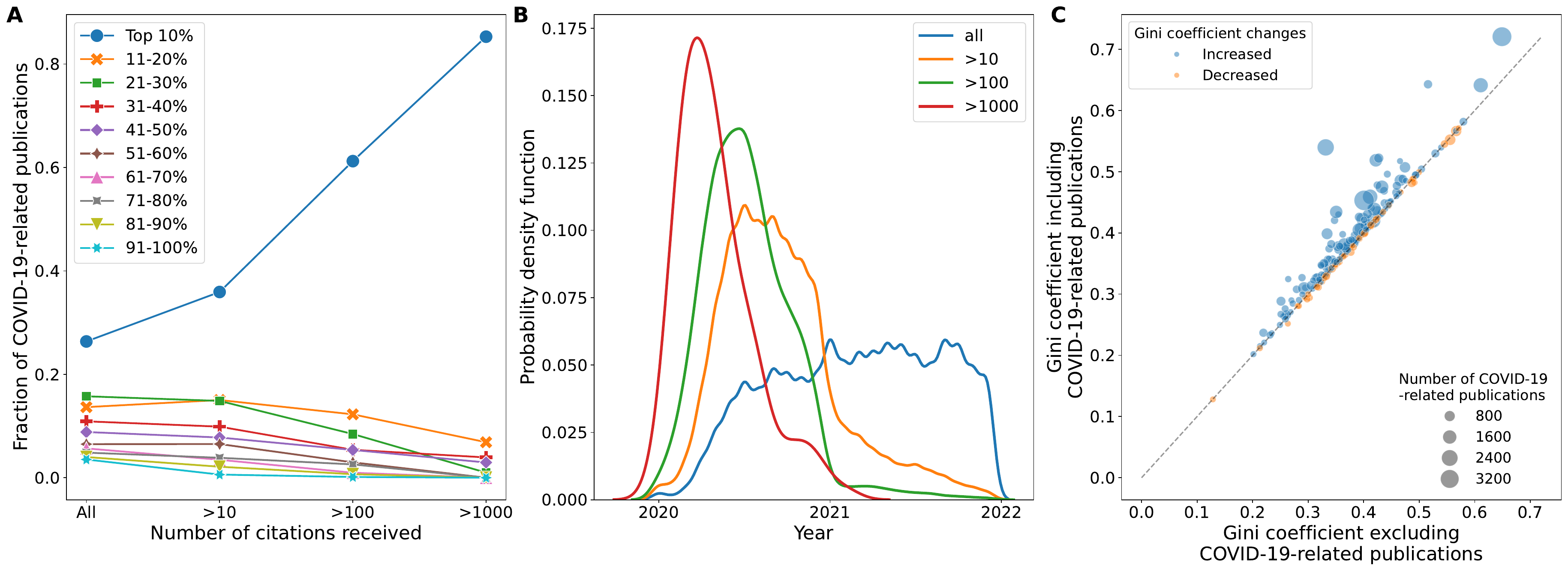}
    \caption{\textbf{Distribution of COVID-19-related publications and their disparities.}
    \textbf{A} Distribution of COVID-19-related research by journal ranking. The journal ranking is computed within the research category.
    \textbf{B} Distribution of COVID-19-related publications by publication date.
    \textbf{C} Plot of the Gini coefficient of the IF distribution by JCR category. Each dot represents a JCR category. The Gini coefficient is computed using the IF distribution of journals in a particular category including and excluding COVID-19-related publications. Blue (orange) dots indicate an increase (decrease) in the Gini coefficient by publishing COVID-19-related research. The size of the dots is proportional to the number of COVID-19-related studies published in the category.
    }
    \label{fig:fig4}
\end{figure}

To support the Matthew effect, we compared the distribution of highly-cited COVID-19-related publications in terms of the journals' IFs and published dates. First, we found that more COVID-19-related publications were published in prestigious journals with a high IF than in other journals (Fig.~\ref{fig:fig4}A). While the top 10\% ranked journals published 26.3\% (51,976) of all COVID-19-related publications from 2019 to 2021, the bottom 90\% to 100\% ranked journals published only 3.5\% (6,977) in the same period. We also observe that the share of COVID-19-related publications decreases as the journal ranking falls (Fig.~\ref{fig:fig4}A). Moreover, the proportion of highly cited papers exacerbates the disparity. 84.3\% (86) of the 102 papers with over 1000 citations were published by the top 10\% journals, while journals ranked 10 to 20 \% published only eight of these studies. No papers with over 1000 citations were published in the bottom 50\% journals.

Citation is a stochastic multiplicative process, whereby papers with a higher citation count are more likely to receive additional citations. Even with similar content between papers, those published in a more prominent location are more likely to be cited~\parencite{lariviere2010impact}. In addition, earlier works may receive more citations because citations are cumulative. Indeed, we find that the majority of highly cited COVID-19-related publications were published during the early stage of the pandemic (early 2020), as shown in Fig.~\ref{fig:fig4}B. The number of COVID-19-related publications gradually increased as the pandemic progressed (see the blue line in Fig.~\ref{fig:fig4}B). Despite this, papers with higher numbers of citations were generally published earlier. In conjunction with the finding that highly cited studies were likely to be published in prestigious journals, we may conclude that COVID-19-related studies were originally introduced in high IF journals, and that lower IF journals then cited the previous publications from high IF journals.

% Rank 변화
This growing pattern of citations could worsen the polarization of academic journals. Publication of COVID-19-related research gave a significantly greater benefit to journals with higher IFs than to those with lower IFs. Consequently, the relative position (rank) of the majority of journals shows only minor changes, although the overall IF of all journals tended to increase during the pandemic. Only 39.0\% of journals that published COVID-19-related publications moved to a higher rank by including COVID-19-related research in one of their subject categories; among them, only 51.8\% changed their IF quantile to a higher one (Fig.~\ref{fig:sm_rank_change}). The other 61.0\% of journals maintained or decreased their ranking. In addition, significant increases or decreases in the ranks of journals were rarely observed (Fig.~\ref{fig:sm_rank_change}). The majority (90\%) of the top 10\% ranked journals maintained their position regardless of COVID-19 research, while the other 10\% fell into the 10\% to 20\% group.

In summary, i) the IF of journals increased overall by publishing COVID-19-related research, ii) journals with higher IFs received greater benefits by publishing COVID-19-related research, and iii) the relative ranks of journals did not change significantly from publishing COVID-19-related research. These findings lead to an interesting question: Did the publication of COVID-19-related research actually increase the polarization of journals? To answer this, we applied the Gini coefficient~\parencite{gini1912variabilita}, a well-known measure of income inequality, to the distribution of journal IFs. In our investigation, the Gini coefficient measures the distribution of citations across journals within a JCR category, ranging from $0$ for the lowest heterogeneity (when all journals receive the same average number of citations) to $1$ for the highest heterogeneity (when only a single journal receives all citations). The trend illustrated by the difference in the Gini coefficient as a function of the number of COVID-19-related publications (see Figs.~\ref{fig:fig4} and \ref{fig:sm_gini}) implies that the disparity in the number of citations between journals increases as the number of COVID-19-related publications published increases (Table~\ref{tab:sm_corr}). We also found that the inequality increased as more proportion of papers and journals within each JCR category published COVID-19-related publications (Spearman $\rho = 0.596$ and $0.436$, respectively) while the number of journals in the field was not correlated (Spearman $\rho = 0.164$). For instance, the \textit{Infectious diseases} field increased the Gini coefficient 0.208 (from 0.332 to 0.540) by publishing 2,589 COVID-19-related publications. Of all 96 journals in the field, 88 journals published COVID-19-related publications. In conclusion, based on the present snapshot of the Web of Science (WOS) dataset, we found that the general pattern of heterogeneity, or polarization, among journals rises as the number of published COVID-19-related publications increases.

\section{Discussion}

From the outset of the global COVID-19 pandemic, many scholars pursued the topic and published a massive number of studies in an unprecedentedly short period. We discovered a trend that, as a result of the intensive publication, COVID-19-related publications acquired more citations than papers in other domains, which reflects its considerable attention in academia. We uncovered two significant consequences that may have led to a more severe polarization of journals in terms of citations. First, 84\% of journals that published COVID-19-related publications in 2019 and 2020 increased their impact factors. Second, prestigious journals were more likely to publish highly cited COVID-19-related publications than other journals (Fig.~\ref{fig:fig3}).

Nonetheless, we demonstrated that publishing a large number of COVID-19-related publications did not immediately boost a journal's IF. Increasing numbers of COVID-19-related publications published in a journal tended to diminish the citation impact of a single COVID-19-related publication. In addition, we found that prestigious journals with a high IF gained more benefit (increased IF) from publishing COVID-19-related research, and also that the publications receiving the highest number of citations were predominantly published in prestige journals during the early stages of the pandemic. Note that our results should not be confused with the causal relationship between their high citation counts, published journals (high prestige), and time period (early pandemic stage). Given that not all COVID-19-related publications increased their journal's IF, one may assume that prestige journals simply have accepted and published more significant research. However, considering that some papers published in prestige journals were ultimately retracted \parencite{mehra2020retraction_a, mehra2020retraction_b}, the high number of citations given to these journals would be a result of other factors rather than the significance of the works.

% limitation
As we could not explicitly assess the quality of each paper due to the scale of the dataset, it is unclear which of the two aforementioned characteristics (quality or visibility) has a larger impact on the current disparity in benefit from publishing COVID-19-related research. The increase in citation inequality during the COVID-19 pandemic may have been driven by online visibility due to the limited communication channels by restricting offline gatherings. To test this hypothesis, a study of the correlation between citation and metrics that reflect online visibility, such as altimetric, during the pandemic may provide a better understanding of the citation dynamics, but we leave this for future study. We believe that a more in-depth investigation of the relationship between research quality (or significance) and citations may be necessary to increase the impact of our findings. Also, a more detailed understanding of such correlation should form the basis of explaining complex citation dynamics, yet we suggest this as a topic for future research. Moreover, since our analysis does not directly address the motivations behind individual citations, more investigation is needed to determine how the observed results can be explained from a general citation dynamics perspective beyond COVID-19-related research. For instance, we suggest survey studies of citation motivation for the highly-cited but eventually retracted papers will be promising to better understand citation dynamics.

% conclusion
Despite its limitations, this study can provide important insights into citation dynamics and its effects on global events. Because of the rich-get-richer nature of citations, papers published in prestigious journals tend to receive more citations. As the relative ranking of the journals did not change significantly despite the increase in the overall IFs of journals publishing COVID-19-related research, fluctuations in IF may not well reflect the actual impact of academic publications. This effect predominantly benefited well-established journals, while other journals did not experience benefits to the same extent (Fig.~\ref{fig:fig4}). Our research indicates that IFs are vulnerable to external events. The majority of the recent IF changes are attributable to citations of COVID-19-related publications; consequently, after the pandemic is over, the majority of the journals may revert to their pre-pandemic IF levels. It is challenging to evaluate academic journals or other participants (researchers, institutions, etc.) using basic statistics because doing so reflects only a portion of actual scientific achievements. Therefore, the simplified metrics employed by some governments \parencite{van2012intended} should be accompanied by a comprehensive and qualitative analysis of journals and individual papers for assessment.

The use of quantitative indicators such as the IF metric has been under debate. The San Francisco Declaration on Research Assessment (DORA), which serves as the starting point for these discussions, explicitly states that the use of journal-based measures (such as IFs) should be avoided to act as a proxy for the quality of individual research publications, to evaluate the contributions of an individual scientist, or to make hiring, promotion, or funding choices. In practice, however, many funders and institutions employ journal-based measures or the number of citations as markers for evaluation rather than assessing the quality of individual papers. The polarization of citations observed in this study demonstrates the inherent hazard of such indicators. The IF metric is not a stable index against external shocks; it might fluctuate temporarily and then revert following external factors. Along with the other well-known limitations of IF, such as skewed citation distributions within journals~\parencite{van2005characteristics,bornmann2017skewness}, the vulnerability of the IF metric as we found here indicates that it is increasingly inappropriate to consider journal IF as a proxy for an individual paper's quality. 

During the current pandemic, the rapid release of COVID-19-related works resulted in less-qualified academic outputs to the public, leading to the retraction of many publications~\parencite{quinn2021following,el2021publications}. Unfortunately, this issue happened not only in journals with a reputation for a weak review process or low publishing difficulty but also in prestigious journals that are widely respected. Worse still, these retracted works earned a substantial number of citations and extensive media attention~\parencite {khan2022bibliometric}. The general public may assume that papers published in academic journals are trustworthy and may likewise trust secondary sources such as scientific news reporting the results of academic findings. In the current context of appraising science and technology, there is a chance that content published in journals with strong indicators will be considered more reputable. Scientists must inform the public that citation measures and journals are not equivalent to the quality of individual research publications. In other words, the number of citations should not be the defining characteristic of quality research. The contemporary ecosystem of research and technology is seemingly supported by scientists' mutual trust and goodwill, and the public may view the scientific community's findings with a similar level of confidence. Combined with the stability issue of the IF metric identified in this study, shouldn't the current practice of over-reliance on citation indices be discontinued so as not to break this chain of trust? For this reason, we believe that responsible action based on actual societal influence is essential for all members of academia, as opposed to merely producing popular research to boost citation impact and one's professional reputation.

\section*{Acknowledgement}
We appreciate the anonymous reviewers for their dedication, which brought significant improvement to the paper. This research was supported by the MSIT (Ministry of Science and ICT), Republic of Korea, under the Innovative Human Resource Development for Local Intellectualization support program (IITP-2022-RS-2022-00156360) supervised by the IITP (Institute for Information \& Communications Technology Planning \& Evaluation). This work was also supported by the National Research Foundation of Korea (NRF) funded by the Korean government (grant No. NRF-2022R1C1C2004277 (T.Y.) and 2022R1A2C1091324 (J.Y.)). The Korea Institute of Science and Technology Information (KISTI) also supported this research with grant No. K-23-L03-C01 (J.Y.L., J.P.) and by providing KREONET, a high-speed Internet connection. The funders had no role in the study design, data collection and analysis, decision to publish, or preparation of the manuscript.

\section*{Ethics declarations}
\subsection*{Competing interests}
The authors declare no competing interests.

\printbibliography

\clearpage

\appendix

\nolinenumbers

\begin{center}
    \item {\fontsize{14}{0}\selectfont \textbf{Supplementary Information for}}
    \item{\fontsize{14}{0}\selectfont Auditing citation polarization during the early COVID-19 pandemic}
    \item{\fontsize{10}{0}\selectfont Taekho You,  June Young Lee, Jinseo Park$^{*}$, Jinhyuk Yun$^{*}$}
    \item{\fontsize{10}{0}\selectfont $^*$Corresponding author(s). Email: jayoujin@kisti.re.kr, jinhyuk.yun@ssu.ac.kr}
\end{center}

%%%%%%%%%% Merge with supplementary Information %%%%%%%%%%
%%%%%%%%%% Prefix a "S" to all equations, figures, tables and reset the counter %%%%%%%%%%
\setcounter{equation}{0}
\setcounter{figure}{0}
\setcounter{table}{0}
\setcounter{page}{1}
\setcounter{section}{0}

\makeatletter
\renewcommand{\thesection}{Section S\arabic{section}}
\renewcommand{\theequation}{S\arabic{equation}}
\renewcommand{\thefigure}{S\arabic{figure}}
\renewcommand{\figurename}{\textbf{Figure}}
\renewcommand{\thetable}{S\arabic{table}}
\renewcommand{\tablename}{\textbf{Table}}

\noindent\textbf{This PDF file includes:}\\

\begin{figure*}
    \centering
    \includegraphics[width=\linewidth]{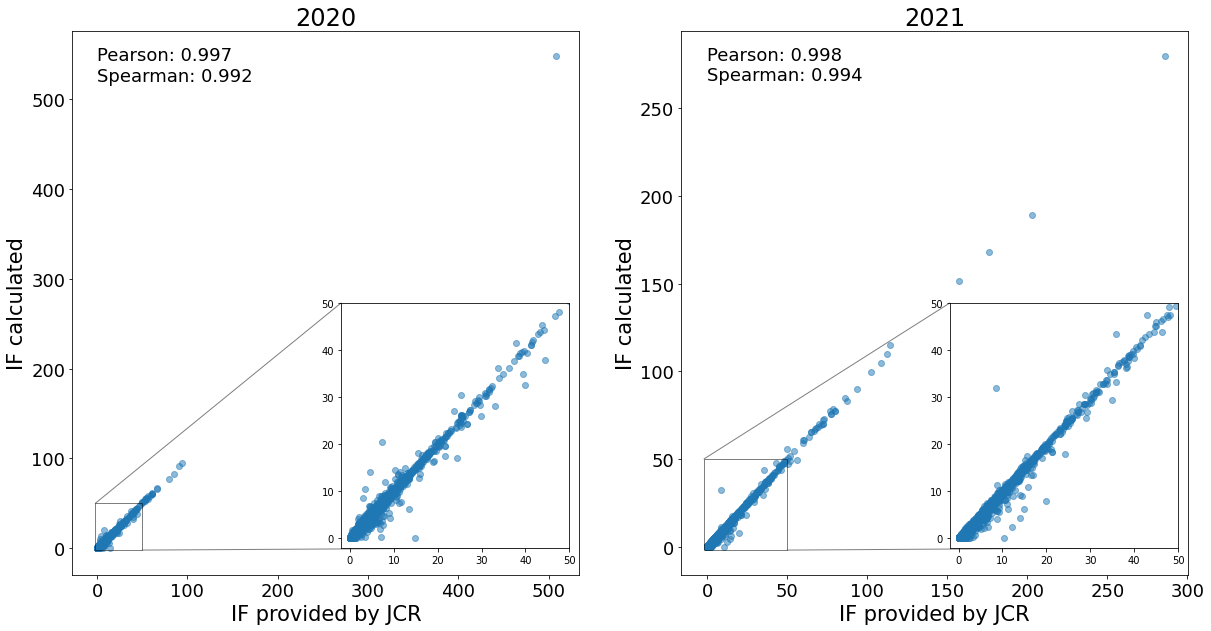}
    \caption{\textbf{Correlation between the IFs provided by JCR and those calculated in this paper.} The Pearson correlation is 0.997 and 0.998 for 2020 and 2021. The Spearman correlation is 0.992 and 0.994 for 2020 and 2021. Insets show the lower part that ranges from 0 to 50.}
    \label{fig:sm_IF}
\end{figure*}

\begin{figure*}
    \centering
    \includegraphics[width=0.8\linewidth]{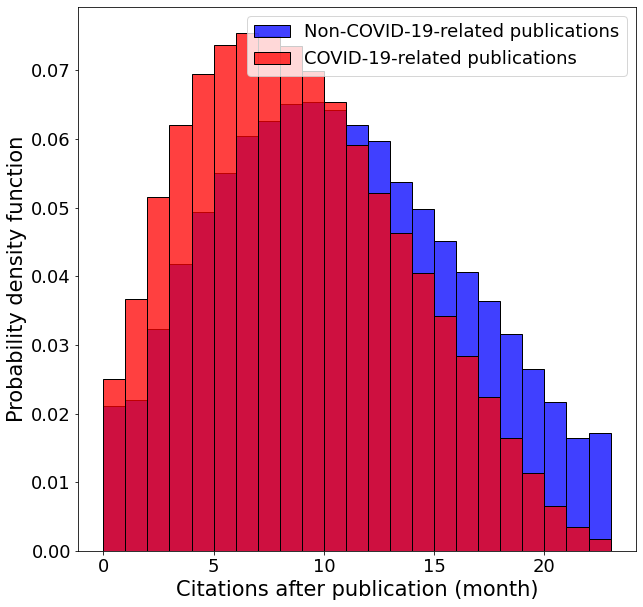}
    \caption{\textbf{Probability density function of the citation time difference between the publication and the citation month of the papers.} For the plot, the COVID-19-related and non-COVID-19-related publications published in 2020 and 2021 were used.}
    \label{fig:sm_cite_month}
\end{figure*}

\begin{figure*}
    \centering
    \includegraphics[width=\linewidth]{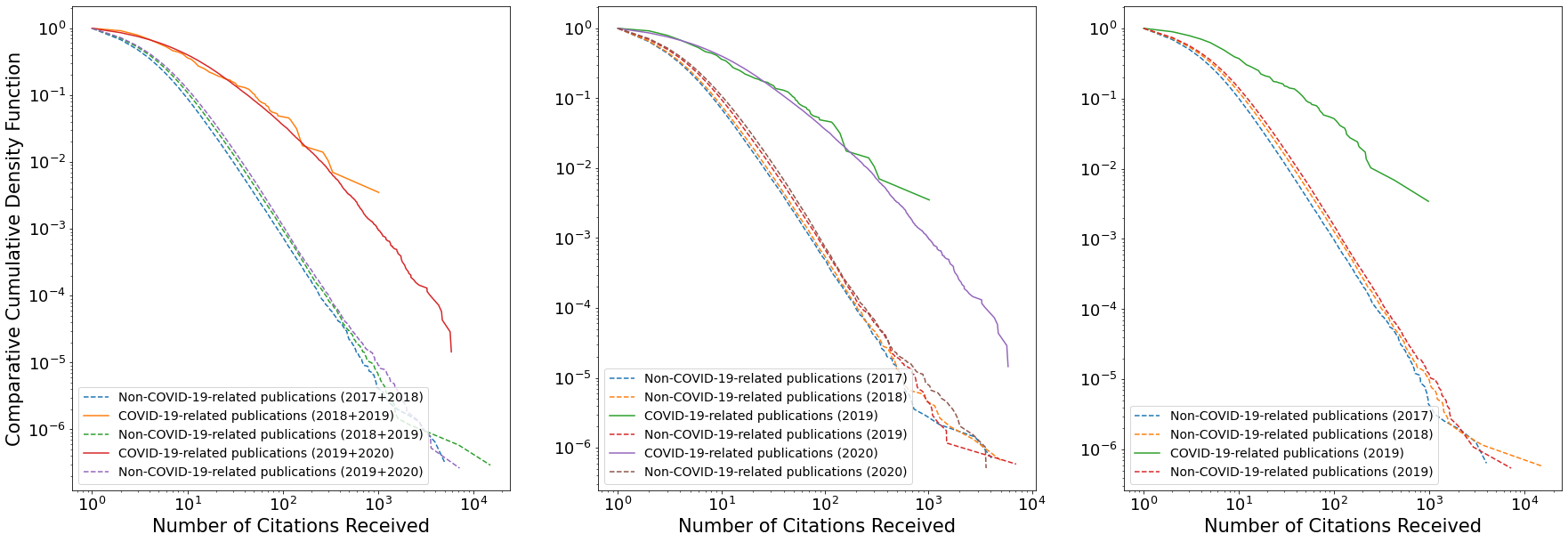}
    \caption{\textbf{Citation distribution of COVID-19-related and non-COVID-19-related publications.} The citation distribution that can contribute to the annual IF calculation (left) is the same distribution as in Fig.~\ref{fig:fig1}A. The separated citation distributions in one-year (middle) and two-year (right) time gaps show the same pattern. Both plots show that COVID-19-related publications have a heavier-tailed distribution.}
    \label{fig:sm_citation_distribution}
\end{figure*}

\begin{figure*}
    \centering
    \includegraphics[width=0.8\linewidth]{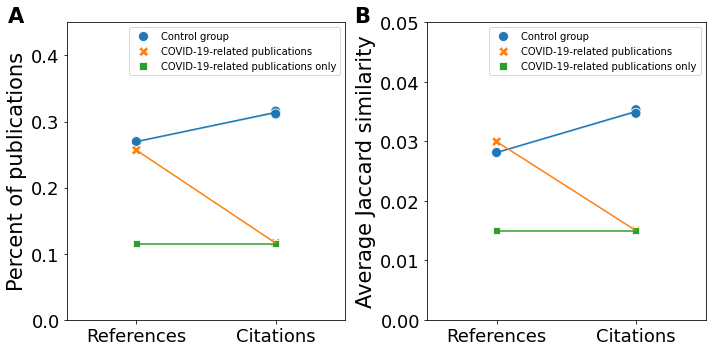}
    \caption{\textbf{Keyword analysis of COVID-19-related publications and its citations and references.} Same number of papers published in the same field and year of COVID-19-related publications are considered as a control group. The control group papers are randomly selected 20 times. \textbf{A} Percent of references or citations that share at least one keyword from the target paper. 
    \textbf{B} Average keyword Jaccard similarity of the target paper and a reference or a citation.}
    \label{fig:sm_keyword}
\end{figure*}

\begin{figure*}
    \centering
    \includegraphics[width=0.8\linewidth]{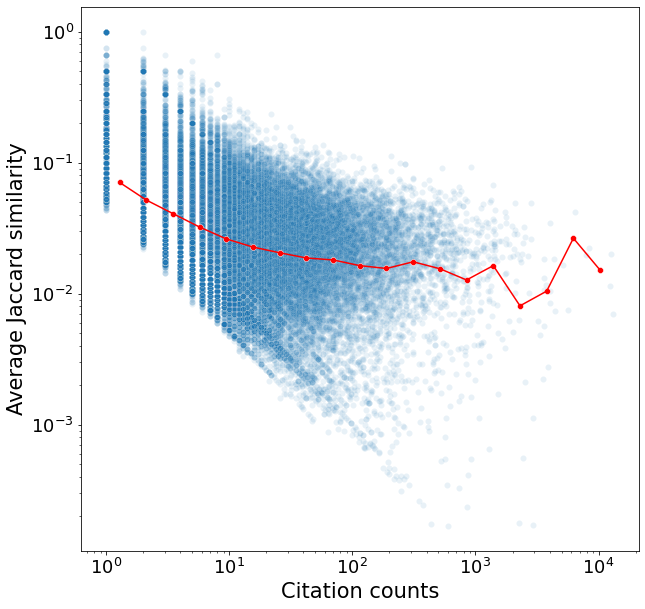}
    \caption{\textbf{Average Jaccard similarity of keywords between COVID-19-related publications and its citations.} Citation counts are calculated only for citations from other COVID-19-related publications. Red dots represent average values. Both Pearson correlation and Spearman rank correlation are $-0.521$.}
    \label{fig:sm_keyword_citation}
\end{figure*}

\begin{figure*}
    \centering
    \includegraphics[width=0.8\linewidth]{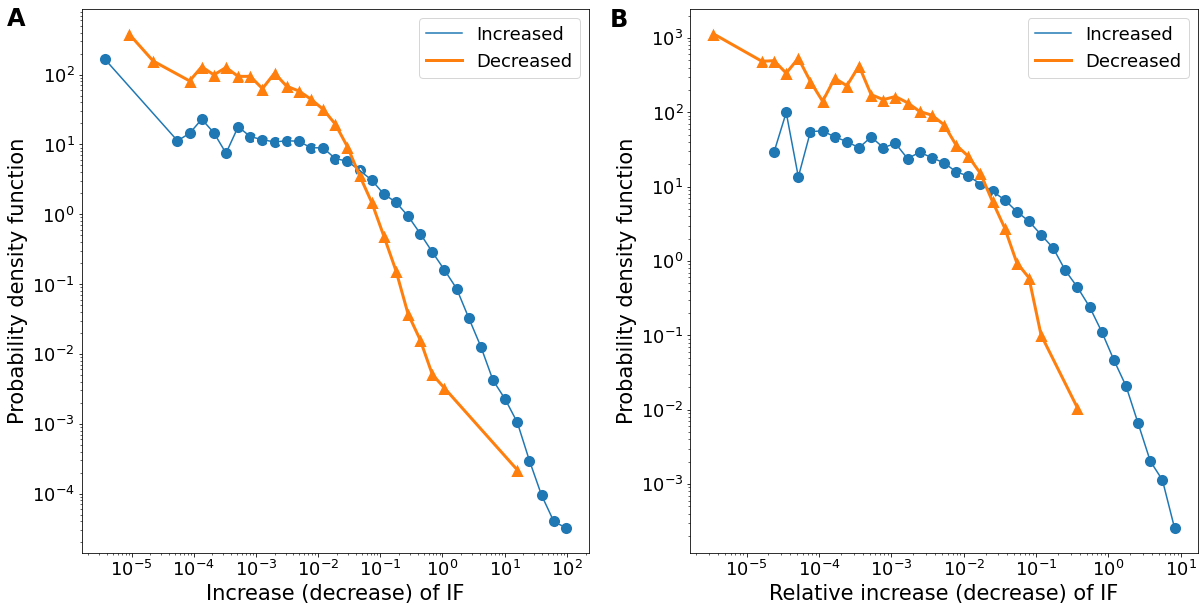}
    \caption{\textbf{Probability density function of journals with IFs that increased or decreased by publishing COVID-19-related publications.} The IFs of 763 journals (16\%) decreased while the IFs of 4004 journals (84\%) increased.
    \textbf{A} The pdf for the increase in IF.
    \textbf{B} The pdf for the increase in IF divided by the IF excluding COVID-19-related publications.}
    \label{fig:sm_if_change}
\end{figure*}

\begin{figure*}
    \centering
    \includegraphics[width=0.8\linewidth]{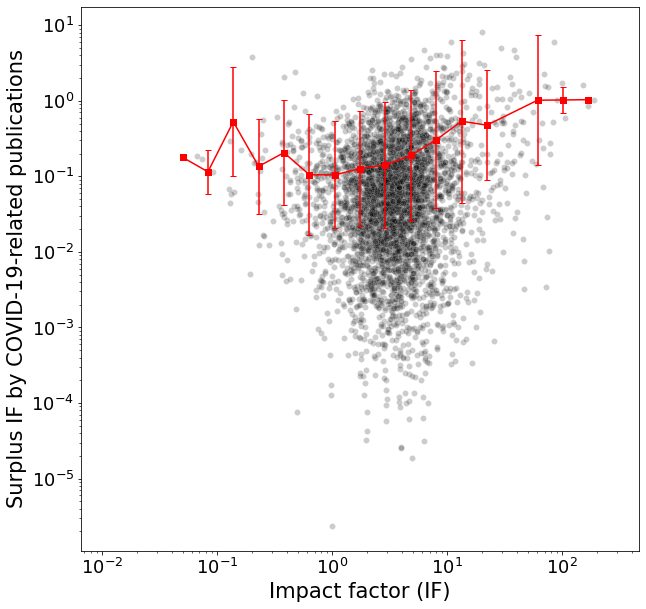}
    \caption{\textbf{Relative IF increase by publishing COVID-19-related publications relative to the journal's IF.} The extent of IF increase is divided by the IF excluding COVID-19-related publications. The red squares and error bars respectively show the average value and the standard deviation of the increased IF in log-scale.}
    \label{fig:sm_IF_increase_relative}
\end{figure*}

\begin{figure*}
    \centering
    \includegraphics[width=0.8\linewidth]{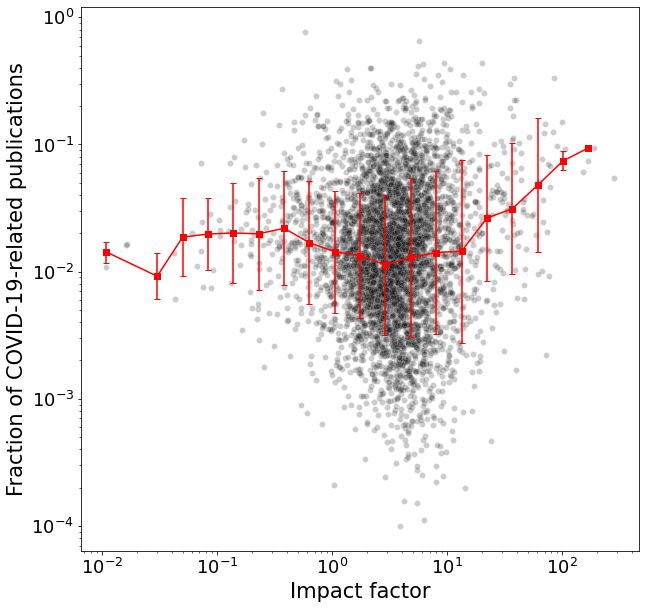}
    \caption{\textbf{Fraction of COVID-19-related publications published in journals by the journal IFs.} The red squares and error bars respectively show the average value and the standard deviation of the fraction of COVID-19-related publications in log-scale. The Spearman rank correlation is $0.008$.}
    \label{fig:sm_covid_IF_correlation}
\end{figure*}

\begin{figure*}
    \centering
    \includegraphics[width=0.8\linewidth]{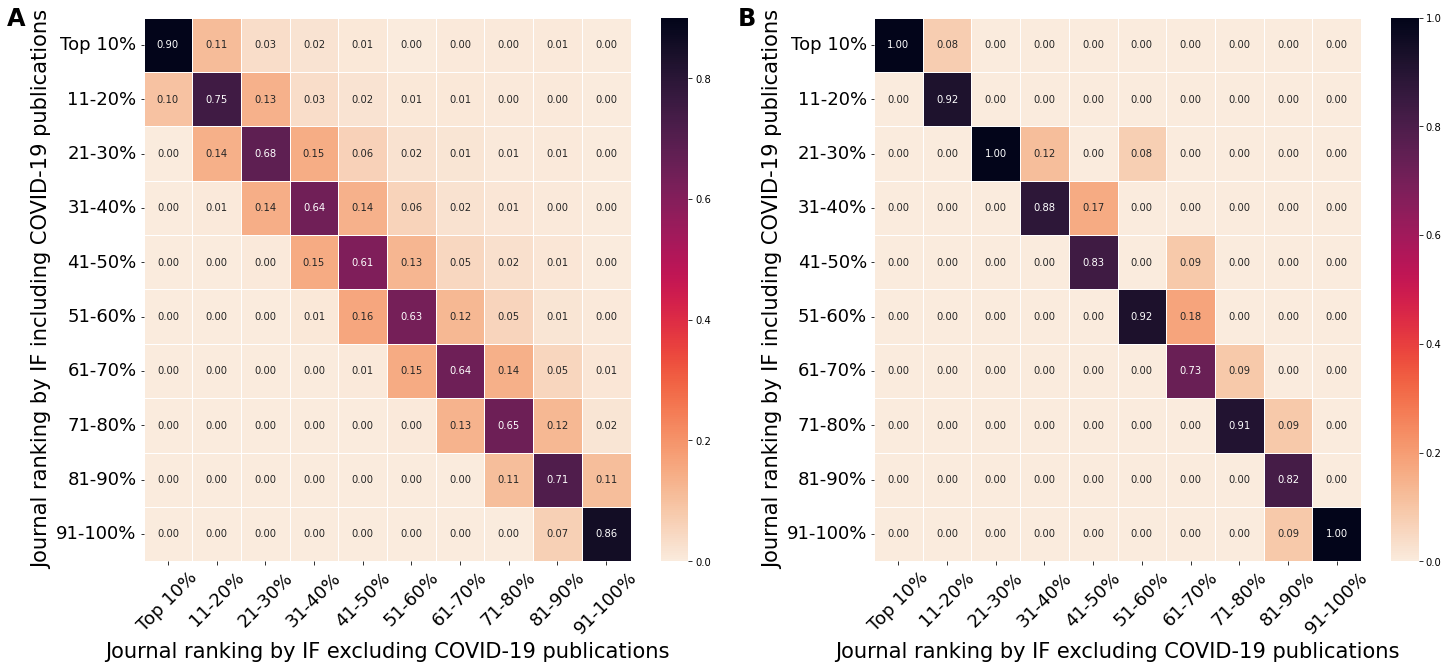}
    \caption{\textbf{Change in journal ranking by publishing COVID-19-related publications.}  The values show the change rate between ranking groups by publishing COVID-19-related publications. Both Pearson and Spearman rank correlations of the IF ranks are 0.99.
    \textbf{A} Rank change of all journals that published COVID-19-related publications in their category.
    \textbf{B} Rank change of the journals that published COVID-19-related publications, where less than 10\% of the journals published COVID-19-related publications in their category. }
    \label{fig:sm_rank_change}
\end{figure*}

\begin{figure*}
    \centering
    \includegraphics[width=0.8\linewidth]{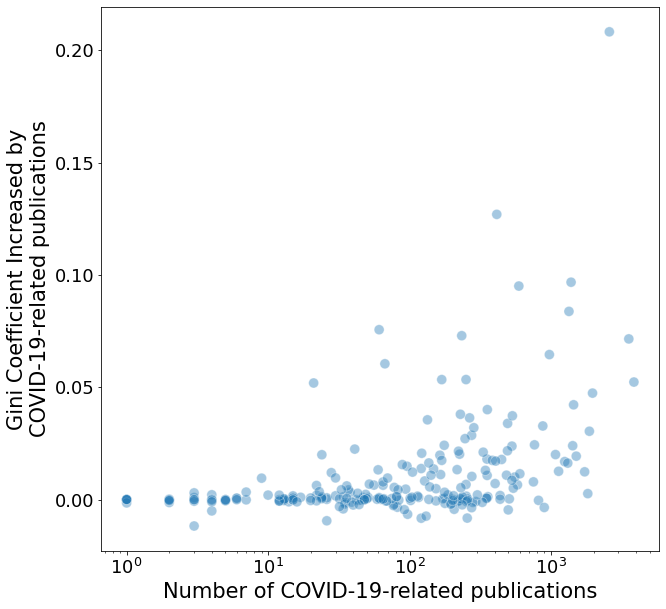}
    \caption{\textbf{Change in the Gini coefficients of the JCR categories by the number of published COVID-19-related publications.} A positive correlation between the number of COVID-19-related publications and the changes in the Gini coefficients of the categories is observed (Spearman $\rho=0.480$).}
    \label{fig:sm_gini}
\end{figure*}

Supplementary tables \ref{tab:field} to \ref{tab:sm_corr}\\
Figures. \ref{fig:sm_IF} to \ref{fig:sm_gini}\\

\newpage

\newpage

\onecolumn
\begin{sidewaystable}
    %\sidewaystablefn% 
    \centering
    \scalebox{0.77}{
    \begin{minipage}{1.15\textheight}
    \caption{\textbf{Relative ratio of surplus impact factor (IF) from publishing COVID-19-related publications by journal category classified by JCR.} 
    The bold numbers represent the highest ratio of surplus IF in each category.
    }
    \label{tab:field}
    \begin{tabular*}{1.15\textheight}{l|rrrrrrrrrr}
        Journal Category & Top 10\% & 11--20\% & 21--30\% & 31--40\% & 41--50\% & 51--60\% & 61--70\% & 71--80\% & 81--90\% & 91--100\% \\
        \midrule
Agricultural Science & \textbf{1.664} & 1.098 & 1.111 & 1.040 & 1.002 & 1.029 & 1.078 & 1.051 & 1.039 & 1.062 \\
Arts \& Humanities, Interdisciplinary & 1.209 & \textbf{1.399} & 1.146 & 1.046 & 0.999 & 1.086 & 1.165 & 0.995 & 0.990 & 0.982 \\
Biology \& Biochemistry & \textbf{1.286} & 1.103 & 1.101 & 1.090 & 1.064 & 1.067 & 1.115 & 1.062 & 1.066 & 1.081 \\
Chemistry & \textbf{1.104} & 1.043 & 1.040 & 1.042 & 1.038 & 1.055 & 1.063 & 1.026 & 1.046 & 1.109 \\
Clinical Medicine & \textbf{1.508} & 1.251 & 1.189 & 1.162 & 1.152 & 1.127 & 1.142 & 1.106 & 1.102 & 1.111 \\
Computer Science & 1.112 & 1.118 & \textbf{1.125} & 1.079 & 1.061 & 1.038 & 1.090 & 1.041 & 1.071 & 1.053 \\
Economics \& Business & \textbf{1.423} & 1.161 & 1.180 & 1.115 & 1.105 & 1.088 & 1.103 & 1.075 & 1.078 & 1.024 \\
Engineering & 1.044 & \textbf{1.109} & 1.041 & 1.048 & 1.029 & 1.053 & 1.027 & 1.026 & 1.037 & 1.146 \\
Environment/Ecology & \textbf{1.388} & 1.209 & 1.201 & 1.135 & 1.129 & 1.112 & 1.126 & 1.106 & 1.074 & 1.121 \\
Geosciences & 1.024 & 1.012 & \textbf{1.037} & 1.018 & 1.018 & 1.002 & 1.021 & 1.033 & 1.041 & 1.001 \\
History \& Archaeology & 1.162 & \textbf{1.165} & 1.121 & 1.072 & 1.114 & 0.994 & 1.069 & 1.003 & 0.988 & 0.969 \\
Literature \& Language & 1.106 & \textbf{1.175} & 1.111 & 1.205 & 1.026 & 1.136 & 1.110 & 1.050 & 1.019 & 1.045 \\
Material Science & 1.017 & 1.018 & 1.010 & 1.024 & 1.009 & 1.010 & 1.008 & 1.011 & 1.024 & \textbf{1.054} \\
Mathematics & 1.053 & 1.046 & 1.139 & 1.039 & 1.052 & 1.062 & 1.061 & 1.016 & \textbf{1.080} & 1.013 \\
Multidisciplinary & \textbf{1.102} & 1.086 & 1.081 & 1.081 & 1.056 & 1.066 & 1.057 & 1.048 & 1.061 & 1.030 \\
Philosophy \& Religion & \textbf{1.232} & 1.219 & 1.221 & 1.155 & 1.067 & 1.141 & 1.131 & 1.010 & 1.066 & 1.046 \\
Physics & 1.058 & 1.057 & 1.039 & 1.025 & 1.018 & 1.030 & 1.030 & 1.028 & 1.014 & \textbf{1.073} \\
Plant \& Animal Science & \textbf{1.102} & 1.046 & 1.055 & 1.020 & 1.068 & 1.063 & 1.096 & 1.022 & 1.039 & 1.027 \\
Psychiatry/Psychology & \textbf{1.666} & 1.341 & 1.183 & 1.153 & 1.121 & 1.227 & 1.137 & 1.176 & 1.137 & 1.136 \\
Social Sciences, General & \textbf{1.389} & 1.218 & 1.182 & 1.147 & 1.130 & 1.073 & 1.087 & 1.070 & 1.088 & 1.033 \\
Visual \& Performing Arts & 1.088 &
1.095 & 1.069 & 1.109 & 1.049 & 1.138 & 1.038 & 1.041 & 1.025 & \textbf{1.115} \\
        \bottomrule
    \end{tabular*}
    \end{minipage}
    }
\end{sidewaystable}

\begin{table}[]
    \centering
    \caption{Spearman correlation with Gini coefficients and properties of journals in JCR category}
    \scalebox{0.6}{
        \begin{tabular}{c|c|c|c}
         & \makecell{Gini coefficient of IF\\ including COVID-19-related publications} & \makecell{Gini coefficient of IF\\ excluding COVID-19-related publications} & Gini coefficient changes \\
         \midrule
         Number of papers & 0.472 & 0.469 & 0.099 \\
         Number of COVID-19-related publications & 0.312 & 0.203 & 0.480 \\
         \makecell{Proportion of\\COVID-19-related publications} & 0.128 & -0.022 & 0.596\\
         Number of journals & 0.343 & 0.356 & 0.164 \\
         \makecell{Number of journals that\\ published COVID-19-related publications} & 0.322 & 0.242 & 0.381 \\
         \makecell{Proportion of journals that\\ published COVID-19-related publications} & 0.093 & -0.029 & 0.436
    \end{tabular}
    }
    \label{tab:sm_corr}
\end{table}

\end{document}